\documentclass[superscriptaddress,aps,journal=prl,preprint]{revtex4-2}
\usepackage{graphicx}
\usepackage{epstopdf}

\usepackage{bm,amsmath,amssymb} 
\usepackage{color, soul} 
\usepackage{dcolumn}   
\usepackage{multirow}

\newcommand{\eg}{{\em e.g.,}}
 
\newcommand{\POP}{$\mathcal{P}_{\rm OP}$}
\newcommand{\PIP}{$\mathcal{P}_{\rm IP}$}
\newcommand{\ainse}{$\alpha$-In$_2$Se$_3$}
\newcommand{\binse}{$\beta$-In$_2$Se$_3$}
\newcommand{\bpinse}{$\beta'$-In$_2$Se$_3$}
\newcommand{\Eb}{$E_{\rm b}$}
\newcommand{\WM}{$W_{M}$}
\newcommand{\WD}{$W_{\rm 2D}$}
\newcommand{\aup}{$\alpha_{\rm up}$}
\newcommand{\adn}{$\alpha_{\rm dn}$}
\begin{document}

\title{Competing charge transfer and screening effects in two-dimensional ferroelectric capacitors}

\author{Jiawei Huang$^\dag$}
\affiliation{Key Laboratory for Quantum Materials of Zhejiang Province, Department of Physics, School of Science, Westlake University, Hangzhou, Zhejiang 310030, China}
\author{Changming Ke$^\dag$}
\affiliation{Key Laboratory for Quantum Materials of Zhejiang Province, Department of Physics, School of Science, Westlake University, Hangzhou, Zhejiang 310030, China}
\author{Zhuang Qian}
\affiliation{Key Laboratory for Quantum Materials of Zhejiang Province, Department of Physics, School of Science, Westlake University, Hangzhou, Zhejiang 310030, China}
\author{Shi Liu}
\email{liushi@westlake.edu.cn}
\affiliation{Key Laboratory for Quantum Materials of Zhejiang Province, Department of Physics, School of Science, Westlake University, Hangzhou, Zhejiang 310030, China}

\date{\today}

\begin{abstract}{
Two-dimensional (2D) ferroelectrics offer the potential for ultrathin flexible nanoelectronics, typically utilizing a metal-ferroelectric-metal sandwich structure as the functional unit. Electrodes can either contribute free carriers to screen the depolarization field, enhancing nanoscale ferroelectricity, or they can induce charge doping, disrupting the long-range crystalline order. Here, we explore the dual roles of electrodes in 2D ferroelectric capacitors, supported by extensive first-principles calculations covering a range of electrode work functions. 
Our results reveal volcano-type relationships between ferroelectric-electrode binding affinity and work function, which are further unified by a quadratic scaling between the binding energy and the transferred interfacial charge. At the monolayer limit, the charge transfer dictates the ferroelectric stability and switching properties. This is a general characteristic confirmed in various 2D ferroelectrics including \ainse, CuInP$_2$S$_6$, and SnTe. As the ferroelectric layer's thickness increases, the stability of the capacitor evolves from a charge transfer-dominated to a screening-dominated state. The delicate interplay between these two effects will have important implications for the applications of 2D ferroelectric capacitors. 
}

\end{abstract}

\maketitle

\newpage

The discovery of 2D ferroelectrics, specifically those with out-of-plane polarization (\POP) such as $\alpha$-In$_2$Se$_3$~\cite{Ding17p14956,Xiao18p227601} and CuInP$_2$S$_6$~\cite{Liu16p12357,Brehm19p43}, has profoundly reshaped our understanding of ferroelectricity at the nanoscale. Recent experiments show that van der Waals (vdW) stacked bilayers, even if they are composed of nonferroelectric monolayers, can be transformed into ferroelectrics through sliding and twisting operations~\cite{Zheng20p71,Stern21p1462,Rogee22p973,Wu21pe2115703118,Miao22p1158},  thereby greatly expanding the family of 2D ferroelectrics with \POP. These 2D ferroelectrics, benefiting from the uniform atomic thickness, absence of surface dangling bonds, and scalability in the lateral dimension,
are intensely studied as critical components in a wide array of device types, including those used in neuromorphic computing and ultrathin flexible nanoelectronics~\cite{Ielmini18p333,Yu17p1700461}. 
Notably, 2D $\alpha$-In$_2$Se$_3$ stands out for its exceptional carrier mobility, semiconducting bandgap, and robust room-temperature monolayer-limit \POP~even against unscreened depolarization field ($\mathcal{E}_d$) under open-circuit (OC) electrical boundary conditions~\cite{Ding17p14956,Si19p580}. 
Shortly after its discovery in 2017, this 2D ferroelectric was swiftly adopted in the fabrication of ferroelectric semiconductor field effect transistors~\cite{Si19p580} and ferroelectric memristor~\cite{Zhang21p2100609}. 

Most ferroelectric-based electronics, such as ferroelectric random-access memory and ferroelectric tunneling junction~\cite{Wan19p1808606,Garcia14p4289}, employ a two-terminal metal-ferroelectric-metal ($MFM$) sandwich structure in which a ferroelectric is sandwiched between two metal electrodes. It is widely recognized that the interface between metal and ferroelectric layers can have significant impacts on device properties~\cite{Junquera03p506, Xue21p7291,Gerra06p107603,Yang22p1422}. Specifically, the screening of interfacial polarization bound charges by the metallic electrodes is crucial for the stabilization of \POP~normal to the interface. 
It can be reasonably deduced that \POP~in a 2D ferroelectric, intrinsically stable under OC boundary conditions, would be further stabilized when placed between electrodes as the screening could reduce the depolarization field. However, another perspective arises when considering the work functions of the electrodes. 
The 2D ferroelectric in an $MFM$ might experience charge doping due to the charge transfer between the electrodes and the ferroelectric layer~\cite{Xue21p7291,Wang12p247601,Zhang17p1703543,Lin21p7}.
Given the ultrathin nature of 2D ferroelectrics and the sensitivity of spontaneous polarization to charge-doping, a pressing question emerges: Which effect predominates at the nanoscale, and consequently, how does this influence the stability of \POP~considering these two competing effects? 

\begin{figure}[ht]
\centering
\includegraphics[scale=1]{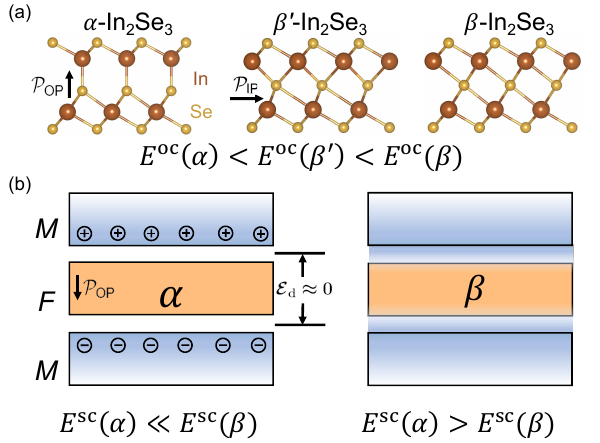}
 \caption{(a) Structures of 2D $\alpha$-In$_2$Se$_3$ with out-of-plane polarization (\POP), $\beta'$-In$_2$Se$_3$ with in-plane polarization (\PIP), and nonpolar $\beta$-In$_2$Se$_3$. An isolated free-standing monolayer is under open-circuit (OC) electrical boundary conditions.
 (b) 
 Illustration of the potential outcomes of an $MFM$ capacitor resulting from the competition between screening and interfacial charge transfer effects. The screening of the depolarization field by the electrodes could further stabilize the $\alpha$ phase under short-circuit (SC) boundary conditions (left). In contrast, electrodes
with low work functions might effectively charge dope the $\alpha$ phase to suppress the ferroelectricity (right).
}
 \label{model}
 \end{figure}

Using In$_2$Se$_3$ as a representative example, here, we elaborate the potential outcomes stemming from the interplay between screening and interfacial charge transfer.
As shown in Fig.~\ref{model}(a), monolayer In$_2$Se$_3$ consists of five atomic planes of triangular lattices in a Se-In-Se-In-Se order, and can crystallize in multiple phases. In monolayer \ainse~(space group $P3m1$), a vertical displacement of the middle Se layer along the $z$-axis leads to \POP; in contrast, the $\beta'$ phase exhibits an in-plane polarization (\PIP) due to an in-plane shift of the central Se layer. The high-temperature $\beta$ phase is nonpolar with a space group of $P\bar{3}m1$.
Previous studies based on first-principles density functional theory (DFT) calculations have revealed that the free-standing monolayer \binse~and \bpinse~are higher in energy ($E$) by 0.15 eV and 0.05 eV per 5-atom unit cell (uc), respectively, than the $\alpha$~phase under OC conditions, that is, $E^{\rm{OC}}(\alpha) < E^{\rm{OC}}(\beta') < E^{\rm{OC}}(\beta)$~\cite{Ding17p14956, Wu21p174107}.
This is consistent with a relatively high ferroelectric-paraelectric ($\alpha \to \beta$) phase transition temperature of 700~K as observed in 3-nm-thick \ainse~crystals~\cite{Xiao18p227601}. In an $MFM$ heterostructure incorporating a monolayer \ainse, a potential outcome is that the $\alpha$ phase may be further stabilized relative to the $\beta$ phase under short-circuit (SC) electrical boundary conditions. This stabilization could occur because the free carriers in the capping electrodes might screen the surface polarization bound charges, thereby reducing the depolarization field. Such scenario, where the screening effect is predominant, is denoted as $E^{\rm{SC}}(\alpha) \ll E^{\rm{SC}}(\beta)$ in Fig.~\ref{model}(b). 
Conversely, electrodes with low work functions might effectively dope the monolayer \ainse~with electrons. This could lead to a strong screening of the dipole-dipole interactions, which are crucial for the emergence of ferroelectricity.
Such charge doping likely causes $E^{\rm{SC}}(\alpha) > E^{\rm{SC}}(\beta)$, potentially quenching the ferroelectricity. Considering these conflicting outcomes, comprehending the intricate interplay between screening and interfacial charge transfer becomes essential for advancing and fine-tuning nanoelectronics based on 2D ferroelectrics. 

Here, we address the fundamental question above by analyzing the binding energy ($E_{\rm b}$) between monolayer In$_2$Se$_3$ in different phases and a range of metal electrodes with varying work functions ($W_{M}$), employing first-principles DFT calculations. 
Our results reveal a volcano-type relationship with a flattened top between \Eb~and $W_M$.  A unified relationship between \Eb~and the amount of charge ($Q$) transferred at the interface is established, given by $E_{\rm b}=-\gamma Q^2$ with $\gamma$ as a constant independent of the polar state or phase.
In particular, the nonpolar $\beta$ phase binds more strongly to low-work-function (LW) electrodes compared to the polar $\alpha$ and $\beta'$ phases. Consequently, monolayer In$_2$Se$_3$ favors the nonpolar $\beta$ phase when sandwiched by LW electrodes, whereas it transitions to the polar $\alpha$ phase in the presence of high-work-function (HW) electrodes. 
The substantial impact of interfacial charge transfer on the ferroelectric stability proves to be a prevalent feature for 2D ferroelectrics including CuInP$_2$S$_6$ (CIPS) and SnTe\cite{Liu18p027601}. 
Finally, we observe a crossover from charge-transfer dominance to screening dominance with the increasing thickness of the ferroelectric layer in an $MFM$ structure composed of symmetric LW electrodes and In$_2$Se$_3$. This transition is accompanied by the spontaneous formation of a buffer layer near one electrode, yielding an intrinsically asymmetric $MFM$ capacitor. These insights into the mechanisms governing charge transfer and screening at the nanoscale have important implications for the design of 2D ferroelectric-based ultrathin nanoelectronics.

All DFT calculations are performed using Vienna $ab~initio$ simulation package (VASP)~\cite{Kresse96p11169,Kresse96p15} with the Perdew-Burke-Ernzerhof (PBE) density functional~\cite{Perdew96p3865} and the projector augmented wave (PAW) method~\cite{Blochi94p17953}. Various factors including the interface chemistry and strain conditions due to electrode-ferroelectric lattice mismatch can substantially affect the electronic structure of the metal–ferroelectric interface, making it challenging to disentangle their individual contributions. Here, we select 2D metallic materials with lattice constants close to the ground-state values of \ainse~as electrodes. This approach offers two key advantages. First, the electrode-ferroelectric interface in vdW heterostructures of 2D materials is free from dangling bonds and less affected by lattice mismatch. Second, the extensive variety of 2D metallic materials available in DFT-based database, such as C2DB~\cite{Haastrup18p042002,Gjerding21p044002}, allows for the selection of electrodes with both the desired lattice constants and  a range of work functions. We therefore examine 18 2D electrodes with work functions ranging from 2 to 10 eV (see details in Supplementary Material). In a vdW heterostructure composed of stacked In$_2$Se$_3$ and electrodes, the atomic positions are fully optimized with in-plane lattice constants fixed to the values of monolayer \ainse. All calculations are performed using a plane-wave cutoff energy of 500~eV, a $7\times7\times1$ Monkhorst-Pack $k$-point grid, an energy convergence threshold of $1\times10^{-8}$~eV, and a force convergence threshold of $1\times10^{-7}$~eV/\AA. The vacuum layer along the $z$ direction is thicker than 15 \AA, and the dipole correction is employed to remove the artificial electric field and unphysical dipole-dipole interactions between different periodic images.

\begin{figure}[ht]
\centering
\includegraphics[scale=1]{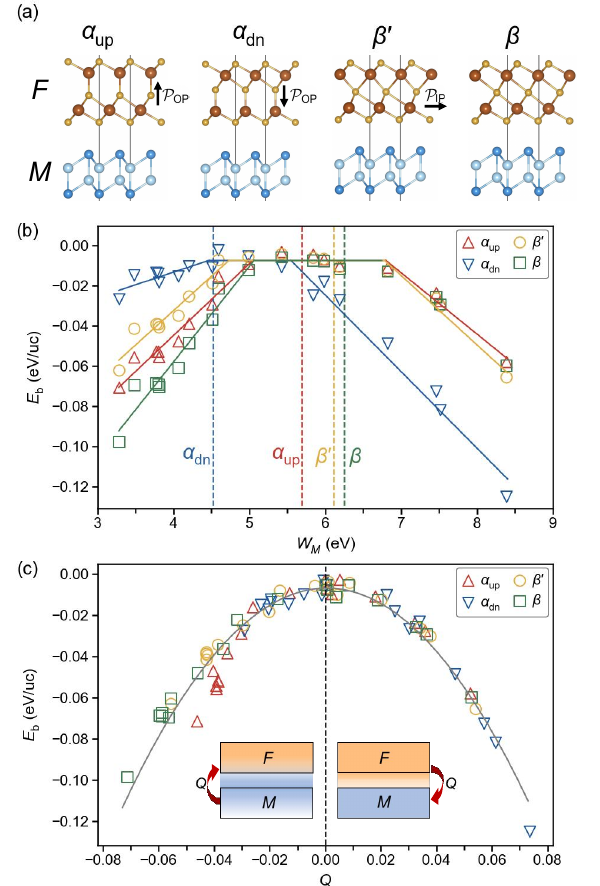}
\caption{(a) Monolayer In$_2$Se$_3$ atop an electrode for binding energy (\Eb) calculations. 
2D metallic materials with lattice constants comparable with \ainse~are chosen as the electrodes. Th specific electrode shown here is Ti$_2$I$_2$ ($W_{M}=3.28$~eV) with I and Ti atoms colored in dark blue and light blue, respectively. (b) Volcano-type relationships between \Eb~and electrode work function $W_M$ for four different bilayer configurations. Vertical dashed lines mark the work functions of In$_2$Se$_3$ in different phases and polar states. (c) \Eb~as a function of the quantify of charge transfer ($Q$) at the interface.  The gray line is the quadratic fit, $E_{\rm b}=-\gamma Q^2$.
}
  \label{chargeTransfer}
 \end{figure}

We start by computing the binding energy of 2D In$_2$Se$_3$ atop a metallic electrode, defined as
$E_{\rm b} = E_{M+\rm 2D} - E_{\rm 2D} - E_{M}$. Here, $E_{\rm 2D}$ and $E_{M}$ denote the energies of the isolated monolayer In$_2$Se$_3$ and the electrode, respectively, while $E_{M+\rm 2D}$ represents the total energy of the bilayer. It is noted that in our calculations of $E_{M+\rm 2D}$, we intentionally fix the interlayer distance at 4.5~\AA, aiming to minimize contributions to \Eb~from factors (\eg~electronegativity of surface atoms) other than the electrode's work function. We consider three phases of In$_2$Se$_3$: $\alpha$, $\beta'$, and $\beta$, all observed experimentally~\cite{Chen23p1077,Zhang19p8004,Zheng22peabo0773,Zheng18peaar7720}. In the case of \ainse-based bilayer heterostructure, there are two configurations to consider.
One has \POP~pointing away from the electrode, and the other has \POP~pointing toward the electrode, labeled as \aup~and \adn~in Fig.~\ref{chargeTransfer}(a), respectively. 

As illustrated in Fig.~\ref{chargeTransfer}(b), the plots of \Eb~versus \WM~curves reveal volcano-type relationships with flattened tops for all bilayer heterostructures. For a specific configuration of In$_2$Se$_3$, the magnitude of \Eb~decreases (becomes less negative) nearly linearly with the increasing \WM~value within the LW region. This is followed by a \WM-insensitive zone (flattened top) when \WM~is comparable to the work function of In$_2$Se$_3$ [\WD, represented by vertical dashed lines in Fig.~\ref{chargeTransfer}(b)]. In the HW region where \WM~is substantially greater than \WD, the binding strength increases with an increasing \WM. Importantly, when the In$_2$Se$_3$ layer is electron-doped by the adjacent LW electrode ($W_{\rm M} < 4.5$~eV), the nonpolar $\beta$ phase exhibits the strongest interaction with the electrode, followed by \aup~and $\beta'$, with \adn~being the weakest. This sequence largely aligns with the order of \WD: $\beta$ has the highest at 6.3 eV, followed by \aup~at 5.6 eV, and \adn~at 4.5~eV. We note that
the work functions of \aup~and \adn~correspond to the work functions of negatively charged and positively charged surfaces of \ainse, respectively, and their difference arises from the built-in depolarization field across the monolayer~\cite{Huang22p1440,Ke21p3387}.
In contrast, in the HW region ($W_{M} >6.5$~eV), \adn~with the lowest work function interacts most strongly with the electrode, whereas the other three types of In$_2$Se$_3$ monolayers have comparable binding energies. These volcano-type relationships cannot be explained by the screening effect. 
Instead, the evident dependence of \Eb~on the difference between \WM~and \WD~highlights the importance of interfacial charge transfer.

We find that the distinct volcano-type relationships presented in Fig.~\ref{chargeTransfer}(b) can be consolidated into a single scaling relationship between \Eb~and the amount of charge ($Q$) transferred at the interface.
A negative $Q$ indicates electron transfer from the electrode to the monolayer In$_2$Se$_3$, typically occurring in the LW region. The hole doping ($Q>0$) of the monolayer In$_2$Se$_3$ is observed near HW electrodes.
Remarkably, as shown in Fig.~\ref{chargeTransfer}(c), the binding energies of 72 bilayer heterostructures (constructed using 18 electrodes and 4 states of 2D In$_2$Se$_3$) are well captured by the quadratic equation $E_{\rm b}=-\gamma Q^2$, with $\gamma=-19.7$ V/e. This quadratic scaling of \Eb~with $Q$ naturally explains the volcano-type relationships between \Eb~and \WM, as $|Q|$ is proportional to $|W_{\rm 2D}-W_{\rm M}|$.
The \WM-insensitive plateaus correspond to weak-charge-transfer regions, where $|Q|<0.05$ e/uc. Results from these model calculations demonstrate conclusively that charge transfer dictates the binding strength between the electrode and the monolayer.

\begin{figure}[ht]
\centering
\includegraphics[scale=0.93]{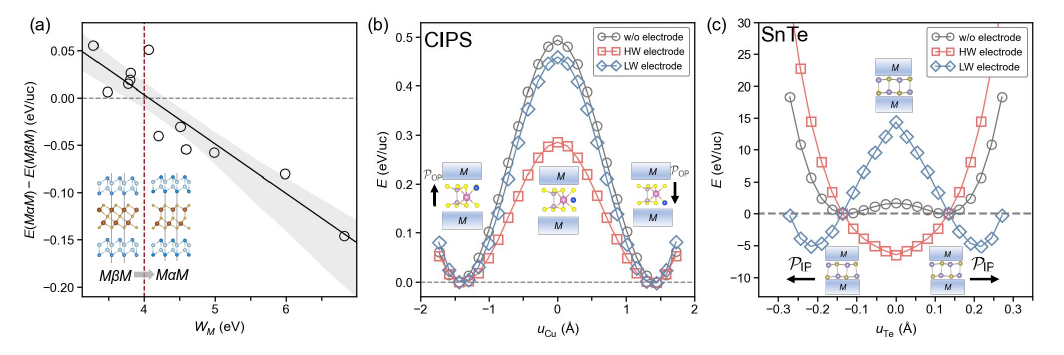}
 \caption{
 (a) Relative thermodynamic stability between $M\alpha M$ and $M\beta M$ as a function of $W_M$. 
 The solid black line represents the linear fit, and the gray area represents the 95\% confidence region.
 The vertical red dashed line marks the critical value of $W_M$, below which the $M\beta M$ capacitor configuration is more stable than $M\alpha M$.   Switching barriers in monolayer (b) CuInP$_2$S$_6$ and (c) SnTe without and with electrodes 
}
  \label{stablePhase}
 \end{figure}

The pronounced interaction between monolayer \binse~and LW electrodes implies that the $\beta$ phase might become energetically favored over the $\alpha$ phase in an $MFM$ trilayer when \Eb~overcompensates for the intrinsic energy difference between these two phases. 
This is confirmed by DFT calculations on realistic capacitor models 
with atomic positions fully relaxed. 
Two trilayer heterostructures are considered, labeled as $M\alpha M$ and $M\beta M$, respectively [see insets in Fig.~\ref{stablePhase}(a)]. Their relative thermodynamic stability is gauged by the energy difference defined as $\Delta E = E({ M\alpha M})-E({ M\beta M})$. It is easy to derive that 
\begin{equation}
\Delta E \propto \left [ E^{\rm OC}(\alpha)-E^{\rm OC}(\beta) \right]+\left [E_{\rm b}(\alpha_{\rm up})+E_{\rm b}(\alpha_{\rm dn})-2E_{\rm b}(\beta) \right ].
\label{eq1}
\end{equation}
Since the freestanding monolayer \ainse~is more stable than \binse, $E^{\rm OC}(\alpha)-E^{\rm OC}(\beta)<0$. However, if $E_{\rm b}(\beta)$ is significantly more negative than $E_{\rm b}(\alpha_{\rm up})$ and $E_{\rm b}(\alpha_{\rm dn})$, the second (positive) term in Eq.~\ref{eq1} overcompensates for the first negative term, reversing the thermodynamic stability between $\alpha$ and $\beta$ phases. Indeed, as Fig.~\ref{stablePhase}(b) illustrates, $M\beta M$ is more stable ($\Delta E > 0$) than $M\alpha M$ for electrodes with $W_{\rm M}<4.0$~eV, whereas $M\alpha M$ becomes the favorable configuration for HW electrodes. 

The significant influence of interfacial charge transfer on ferroelectric stability is a common characteristic of a 2D ferroelectric-based capacitor. Taking CIPS with \POP~and SnTe with \PIP~as examples, we calculate the switching barriers under SC electrical boundary conditions ($\Delta U^{\rm SC}$) with electrodes of LW and HW, respectively (see computational details in Supplementary Material). As shown in Fig.~\ref{stablePhase}(b), the presence of HW electrodes substantially lowers the switching barrier in monolayer CIPS by 0.21 eV/uc. This effect can be understood as follows. Analogous to Eq.~\ref{eq1}, the magnitude of $\Delta U^{\rm SC}$ can be approximated by the energy difference between the ferroelectric (FE) phase and the reference paraelectric (PE) phase in a capacitor, 
\begin{equation}
\Delta U^{\rm SC} \propto  \Delta U^{\rm OC}+ \left [2E_{\rm b}({\rm PE})- E_{\rm b}({\rm FE}_{\rm up})-E_{\rm b}({\rm FE}_{\rm dn}) \right ],
\end{equation}
where $\Delta U^{\rm OC}\approx E^{\rm OC}({\rm PE}) - E^{\rm OC}({\rm FE})$ is the switching barrier in an isolated monolayer. 
Contrary to \binse, the paraelectric phase of CIPS exhibits a stronger binding affinity to HW electrodes than its ferroelectric phase, with $E_{\rm b}({\rm PE})$ being more negative than $E_{\rm b}({\rm FE})$. As a result, HW electrodes enhance the stability of the paraelectric phase relative to the ferroelectric phase, effectively lowering the switching barrier. In comparison, both the ferroelectric and paraelectric phases of CIPS demonstrate similar interaction strengths with LW electrodes, resulting in $\Delta U^{\rm SC}$ being close to $\Delta U^{\rm OC}$. For monolayer SnTe encapsulated by HW electrodes, the effect of charge transfer is similar to that in CIPS: the paraelectric phase is stabilized by HW electrodes more than the ferroelectric phase. Since $\Delta U^{\rm OC}$ is small in SnTe, the in-plane ferroelectricity is then fully suppressed, leading to a nonpolar ground state. 
Interestingly, LW electrodes promote the in-plane ferroelectricity, indicated by an increased switching barrier and a ground state characterized by larger ferroelectric displacements [Fig.~\ref{chargeTransfer}(c)]. These findings indicate that despite the distinct \Eb-$W_M$ relationships in different 2D ferroelectrics,  the significant role of interfacial charge transfer in influencing ferroelectric stability within an $MFM$ capacitor is a common characteristic across these materials.

\begin{figure}[ht]
\centering
\includegraphics[scale=0.8]{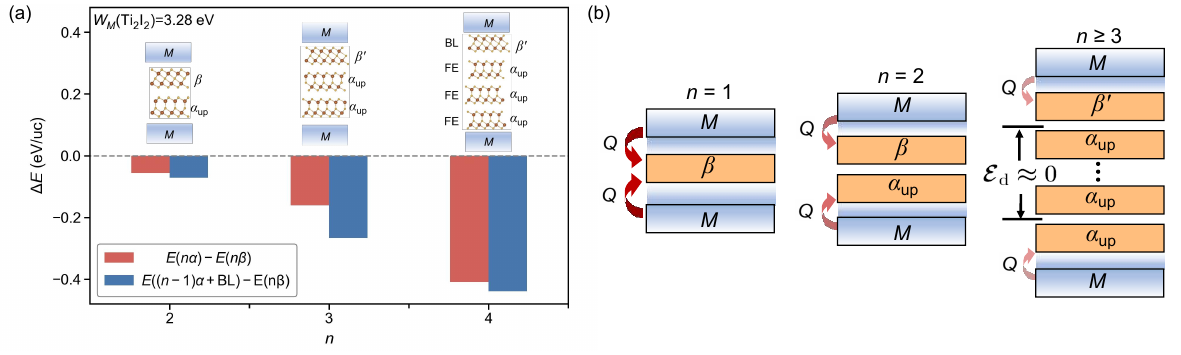}
 \caption{
 (a) Evolution of relative stability of $n$ layers of In$_2$Se$_3$ sandwiched between Ti$_2$I$_2$ electrodes. 
The energy of an $MFM$ consisting of $n$~layers \binse, denoted as $n\beta$, is chosen as the energy reference. The capacitor with $n-1$ layers of \ainse~and one buffer layer (BL) has the lowest energy. 
 The BL adopts the $\beta$ phase at $n=2$ but the $\beta'$ phase for $n\ge 3$. (b) Schematic illustrating the evolution of the capacitor's structure with increasing thickness of the ferroelectric layer, due to the competing charge transfer and screening effects. 
 }
  \label{increaingFE}
 \end{figure}

A pertinent question naturally arises: how does the increasing thickness of the ferroelectric layer influence its stability, particularly considering the competing effects of charge transfer and screening from the electrodes? Since HW electrodes already stabilize \ainse~at the monolayer limit, our discussion below will focus on LW electrodes, represented by Ti$_2$I$_2$ ($W_M=3.28$ eV). We fully optimize the atomic positions of an $MFM$ capacitor consisting of $n$ layers of In$_2$Se$_3$ sandwiched between Ti$_2$I$_2$ electrodes. 
In the case of $n=2$, the bilayer \ainse~becomes more stable than bilayer \binse, as reported in Fig.~\ref{increaingFE}(a). According to Eq.~\ref{eq1}, it is straightforward to derive $\Delta E = E(Mn\alpha M)-E(Mn\beta M) = n\left [ E^{\rm OC}(\alpha)-E^{\rm OC}(\beta) \right]+\left [E_{\rm b}(\alpha_{\rm up})+E_{\rm b}(\alpha_{\rm dn})-2E_{\rm b}(\beta) \right ]$. As the first term, negative and representing bulk contribution, scales with $n$ and the interfacial contribution remains nearly constant, it's unsurprising that the intrinsic thermodynamic stability of the $\alpha$ phase dominates. Interestingly, in the lowest-energy configuration, the bottom layer is in the polar $\alpha$ phase, while the top layer adopting the nonpolar $\beta$ phase functioning as a buffer layer [BL, see insets in Fig.~\ref{increaingFE}(a)]. This suggests that $|E^{\rm OC}(\alpha)-E^{\rm OC}(\beta)| < E_{\rm b}(\alpha_{\rm dn})-E_{\rm b}(\beta)$. For $n\ge 3$, the capacitor with a BL, denoted as $(n-1)\alpha+$BL is always energetically favored over the multilayer \ainse, denoted as $n\alpha$. An intriguing difference from the $n=2$ case is that BL adopts the $\beta'$ phase instead of the $\beta$ phase. Detailed charge analysis reveals that the amount of charge transferred from the top electrode to the adjacent \binse~decreases with increasing thickness (see Supplementary Material). Since \Eb~scales with $Q^2$, the reduced binding affinity at $n=3$ is insufficient to stabilize \binse~over \bpinse. The evolution of the capacitor's structure with increasing thickness of the ferroelectric layer is  illustrated in Fig.~\ref{increaingFE}(b). The spontaneous formation of an asymmetric capacitor with symmetric electrodes could be important for the design of ferroelectric tunneling junctions.

In summary, we have investigated the interplay between interfacial charge transfer and screening effects in 2D ferroelectric capacitors. 
Our extensive model calculations of 72 bilayer heterostructures, involving monolayer In$_2$Se$_3$ in different phases and polar states sandwiched by 2D metallic electrodes, reveal a volcano-type relationship between the binding energy and electrode work function. 
Distinct volcano-type relationships observed among various phases and polar states of In$_2$Se$_3$ can be unified into a robust quadratic correlation, linking the binding energy with the amount of charge transferred at the interface. At the monolayer limit, it is the interfacial charge transfer that determines the relative thermodynamic stability among competing In$_2$Se$_3$ phases in an $MFM$ capacitor. This significant role of interfacial charge transfer in dictating ferroelectric stability and switching barriers is also observed in CuInP$_2$S$_6$ and SnTe, confirming its prevalent importance across 2D ferroelectrics.  With increased ferroelectric layer thickness, we observe a transition from a charge transfer-dominant regime to the screening-dominant regime. In capacitors with multilayer In$_2$Se$_3$, the spontaneous emergence of a buffer layer next to the electrode on one side leads to an inherently asymmetric structure. The insights gained from this study provide useful guidelines for the future design and optimization of next-generation ferroelectric nanoelectronic devices, underlining the significance of interface engineering in these applications.

\begin{acknowledgments}
We acknowledge the supports from National Natural Science Foundation of China (52002335) and Westlake Education Foundation. The computational resource is provided by Westlake HPC Center. 
\end{acknowledgments}

\newpage
\bibliography{SL}

\begin{thebibliography}{36}%
\makeatletter
\providecommand \@ifxundefined [1]{%
 \@ifx{#1\undefined}
}%
\providecommand \@ifnum [1]{%
 \ifnum #1\expandafter \@firstoftwo
 \else \expandafter \@secondoftwo
 \fi
}%
\providecommand \@ifx [1]{%
 \ifx #1\expandafter \@firstoftwo
 \else \expandafter \@secondoftwo
 \fi
}%
\providecommand \natexlab [1]{#1}%
\providecommand \enquote  [1]{``#1''}%
\providecommand \bibnamefont  [1]{#1}%
\providecommand \bibfnamefont [1]{#1}%
\providecommand \citenamefont [1]{#1}%
\providecommand \href@noop [0]{\@secondoftwo}%
\providecommand \href [0]{\begingroup \@sanitize@url \@href}%
\providecommand \@href[1]{\@@startlink{#1}\@@href}%
\providecommand \@@href[1]{\endgroup#1\@@endlink}%
\providecommand \@sanitize@url [0]{\catcode `\\12\catcode `\$12\catcode
  `\&12\catcode `\#12\catcode `\^12\catcode `\_12\catcode `\%12\relax}%
\providecommand \@@startlink[1]{}%
\providecommand \@@endlink[0]{}%
\providecommand \url  [0]{\begingroup\@sanitize@url \@url }%
\providecommand \@url [1]{\endgroup\@href {#1}{\urlprefix }}%
\providecommand \urlprefix  [0]{URL }%
\providecommand \Eprint [0]{\href }%
\providecommand \doibase [0]{https://doi.org/}%
\providecommand \selectlanguage [0]{\@gobble}%
\providecommand \bibinfo  [0]{\@secondoftwo}%
\providecommand \bibfield  [0]{\@secondoftwo}%
\providecommand \translation [1]{[#1]}%
\providecommand \BibitemOpen [0]{}%
\providecommand \bibitemStop [0]{}%
\providecommand \bibitemNoStop [0]{.\EOS\space}%
\providecommand \EOS [0]{\spacefactor3000\relax}%
\providecommand \BibitemShut  [1]{\csname bibitem#1\endcsname}%
\let\auto@bib@innerbib\@empty
\bibitem [{\citenamefont {Ding}\ \emph {et~al.}(2017)\citenamefont {Ding},
  \citenamefont {Zhu}, \citenamefont {Wang}, \citenamefont {Gao}, \citenamefont
  {Xiao}, \citenamefont {Gu}, \citenamefont {Zhang},\ and\ \citenamefont
  {Zhu}}]{Ding17p14956}%
  \BibitemOpen
  \bibfield  {author} {\bibinfo {author} {\bibfnamefont {W.}~\bibnamefont
  {Ding}}, \bibinfo {author} {\bibfnamefont {J.}~\bibnamefont {Zhu}}, \bibinfo
  {author} {\bibfnamefont {Z.}~\bibnamefont {Wang}}, \bibinfo {author}
  {\bibfnamefont {Y.}~\bibnamefont {Gao}}, \bibinfo {author} {\bibfnamefont
  {D.}~\bibnamefont {Xiao}}, \bibinfo {author} {\bibfnamefont {Y.}~\bibnamefont
  {Gu}}, \bibinfo {author} {\bibfnamefont {Z.}~\bibnamefont {Zhang}},\ and\
  \bibinfo {author} {\bibfnamefont {W.}~\bibnamefont {Zhu}},\ }\bibfield
  {title} {\bibinfo {title} {Prediction of intrinsic two-dimensional
  ferroelectrics in {In$_2$Se$_3$} and other {III$_2$-VI$_3$} van der waals
  materials},\ }\href {https://doi.org/10.1038/ncomms14956} {\bibfield
  {journal} {\bibinfo  {journal} {Nat. Commun.}\ }\textbf {\bibinfo {volume}
  {8}},\ \bibinfo {pages} {14956} (\bibinfo {year} {2017})}\BibitemShut
  {NoStop}%
\bibitem [{\citenamefont {Xiao}\ \emph {et~al.}(2018)\citenamefont {Xiao},
  \citenamefont {Zhu}, \citenamefont {Wang}, \citenamefont {Feng},
  \citenamefont {Hu}, \citenamefont {Dasgupta}, \citenamefont {Han},
  \citenamefont {Wang}, \citenamefont {Muller}, \citenamefont {Martin},
  \citenamefont {Hu},\ and\ \citenamefont {Zhang}}]{Xiao18p227601}%
  \BibitemOpen
  \bibfield  {author} {\bibinfo {author} {\bibfnamefont {J.}~\bibnamefont
  {Xiao}}, \bibinfo {author} {\bibfnamefont {H.}~\bibnamefont {Zhu}}, \bibinfo
  {author} {\bibfnamefont {Y.}~\bibnamefont {Wang}}, \bibinfo {author}
  {\bibfnamefont {W.}~\bibnamefont {Feng}}, \bibinfo {author} {\bibfnamefont
  {Y.}~\bibnamefont {Hu}}, \bibinfo {author} {\bibfnamefont {A.}~\bibnamefont
  {Dasgupta}}, \bibinfo {author} {\bibfnamefont {Y.}~\bibnamefont {Han}},
  \bibinfo {author} {\bibfnamefont {Y.}~\bibnamefont {Wang}}, \bibinfo {author}
  {\bibfnamefont {D.~A.}\ \bibnamefont {Muller}}, \bibinfo {author}
  {\bibfnamefont {L.~W.}\ \bibnamefont {Martin}}, \bibinfo {author}
  {\bibfnamefont {P.}~\bibnamefont {Hu}},\ and\ \bibinfo {author}
  {\bibfnamefont {X.}~\bibnamefont {Zhang}},\ }\bibfield  {title} {\bibinfo
  {title} {Intrinsic two-dimensional ferroelectricity with dipole locking},\
  }\href {https://doi.org/10.1103/PhysRevLett.120.227601} {\bibfield  {journal}
  {\bibinfo  {journal} {Phys. Rev. Lett.}\ }\textbf {\bibinfo {volume} {120}},\
  \bibinfo {pages} {227601} (\bibinfo {year} {2018})}\BibitemShut {NoStop}%
\bibitem [{\citenamefont {Liu}\ \emph {et~al.}(2016)\citenamefont {Liu},
  \citenamefont {You}, \citenamefont {Seyler}, \citenamefont {Li},
  \citenamefont {Yu}, \citenamefont {Lin}, \citenamefont {Wang}, \citenamefont
  {Zhou}, \citenamefont {Wang}, \citenamefont {He}, \citenamefont {Pantelides},
  \citenamefont {Zhou}, \citenamefont {Sharma}, \citenamefont {Xu},
  \citenamefont {Ajayan}, \citenamefont {Wang},\ and\ \citenamefont
  {Liu}}]{Liu16p12357}%
  \BibitemOpen
  \bibfield  {author} {\bibinfo {author} {\bibfnamefont {F.}~\bibnamefont
  {Liu}}, \bibinfo {author} {\bibfnamefont {L.}~\bibnamefont {You}}, \bibinfo
  {author} {\bibfnamefont {K.~L.}\ \bibnamefont {Seyler}}, \bibinfo {author}
  {\bibfnamefont {X.}~\bibnamefont {Li}}, \bibinfo {author} {\bibfnamefont
  {P.}~\bibnamefont {Yu}}, \bibinfo {author} {\bibfnamefont {J.}~\bibnamefont
  {Lin}}, \bibinfo {author} {\bibfnamefont {X.}~\bibnamefont {Wang}}, \bibinfo
  {author} {\bibfnamefont {J.}~\bibnamefont {Zhou}}, \bibinfo {author}
  {\bibfnamefont {H.}~\bibnamefont {Wang}}, \bibinfo {author} {\bibfnamefont
  {H.}~\bibnamefont {He}}, \bibinfo {author} {\bibfnamefont {S.~T.}\
  \bibnamefont {Pantelides}}, \bibinfo {author} {\bibfnamefont
  {W.}~\bibnamefont {Zhou}}, \bibinfo {author} {\bibfnamefont {P.}~\bibnamefont
  {Sharma}}, \bibinfo {author} {\bibfnamefont {X.}~\bibnamefont {Xu}}, \bibinfo
  {author} {\bibfnamefont {P.~M.}\ \bibnamefont {Ajayan}}, \bibinfo {author}
  {\bibfnamefont {J.}~\bibnamefont {Wang}},\ and\ \bibinfo {author}
  {\bibfnamefont {Z.}~\bibnamefont {Liu}},\ }\bibfield  {title} {\bibinfo
  {title} {Room-temperature ferroelectricity in {CuInP$_2$S$_6$} ultrathin
  flakes},\ }\href {https://doi.org/10.1038/ncomms12357} {\bibfield  {journal}
  {\bibinfo  {journal} {Nat. Commun.}\ }\textbf {\bibinfo {volume} {7}},\
  \bibinfo {pages} {12357} (\bibinfo {year} {2016})}\BibitemShut {NoStop}%
\bibitem [{\citenamefont {Brehm}\ \emph {et~al.}(2019)\citenamefont {Brehm},
  \citenamefont {Neumayer}, \citenamefont {Tao}, \citenamefont {O'Hara},
  \citenamefont {Chyasnavichus}, \citenamefont {Susner}, \citenamefont
  {McGuire}, \citenamefont {Kalinin}, \citenamefont {Jesse}, \citenamefont
  {Ganesh}, \citenamefont {Pantelides}, \citenamefont {Maksymovych},\ and\
  \citenamefont {Balke}}]{Brehm19p43}%
  \BibitemOpen
  \bibfield  {author} {\bibinfo {author} {\bibfnamefont {J.~A.}\ \bibnamefont
  {Brehm}}, \bibinfo {author} {\bibfnamefont {S.~M.}\ \bibnamefont {Neumayer}},
  \bibinfo {author} {\bibfnamefont {L.}~\bibnamefont {Tao}}, \bibinfo {author}
  {\bibfnamefont {A.}~\bibnamefont {O'Hara}}, \bibinfo {author} {\bibfnamefont
  {M.}~\bibnamefont {Chyasnavichus}}, \bibinfo {author} {\bibfnamefont {M.~A.}\
  \bibnamefont {Susner}}, \bibinfo {author} {\bibfnamefont {M.~A.}\
  \bibnamefont {McGuire}}, \bibinfo {author} {\bibfnamefont {S.~V.}\
  \bibnamefont {Kalinin}}, \bibinfo {author} {\bibfnamefont {S.}~\bibnamefont
  {Jesse}}, \bibinfo {author} {\bibfnamefont {P.}~\bibnamefont {Ganesh}},
  \bibinfo {author} {\bibfnamefont {S.~T.}\ \bibnamefont {Pantelides}},
  \bibinfo {author} {\bibfnamefont {P.}~\bibnamefont {Maksymovych}},\ and\
  \bibinfo {author} {\bibfnamefont {N.}~\bibnamefont {Balke}},\ }\bibfield
  {title} {\bibinfo {title} {Tunable quadruple-well ferroelectric van der
  {Waals} crystals},\ }\href {https://doi.org/10.1038/s41563-019-0532-z}
  {\bibfield  {journal} {\bibinfo  {journal} {Nat. Mater.}\ }\textbf {\bibinfo
  {volume} {19}},\ \bibinfo {pages} {43} (\bibinfo {year} {2019})}\BibitemShut
  {NoStop}%
\bibitem [{\citenamefont {Zheng}\ \emph {et~al.}(2020)\citenamefont {Zheng},
  \citenamefont {Ma}, \citenamefont {Bi}, \citenamefont {de~la Barrera},
  \citenamefont {Liu}, \citenamefont {Mao}, \citenamefont {Zhang},
  \citenamefont {Kiper}, \citenamefont {Watanabe}, \citenamefont {Taniguchi},
  \citenamefont {Kong}, \citenamefont {Tisdale}, \citenamefont {Ashoori},
  \citenamefont {Gedik}, \citenamefont {Fu}, \citenamefont {Xu},\ and\
  \citenamefont {Jarillo-Herrero}}]{Zheng20p71}%
  \BibitemOpen
  \bibfield  {author} {\bibinfo {author} {\bibfnamefont {Z.}~\bibnamefont
  {Zheng}}, \bibinfo {author} {\bibfnamefont {Q.}~\bibnamefont {Ma}}, \bibinfo
  {author} {\bibfnamefont {Z.}~\bibnamefont {Bi}}, \bibinfo {author}
  {\bibfnamefont {S.}~\bibnamefont {de~la Barrera}}, \bibinfo {author}
  {\bibfnamefont {M.-H.}\ \bibnamefont {Liu}}, \bibinfo {author} {\bibfnamefont
  {N.}~\bibnamefont {Mao}}, \bibinfo {author} {\bibfnamefont {Y.}~\bibnamefont
  {Zhang}}, \bibinfo {author} {\bibfnamefont {N.}~\bibnamefont {Kiper}},
  \bibinfo {author} {\bibfnamefont {K.}~\bibnamefont {Watanabe}}, \bibinfo
  {author} {\bibfnamefont {T.}~\bibnamefont {Taniguchi}}, \bibinfo {author}
  {\bibfnamefont {J.}~\bibnamefont {Kong}}, \bibinfo {author} {\bibfnamefont
  {W.~A.}\ \bibnamefont {Tisdale}}, \bibinfo {author} {\bibfnamefont
  {R.}~\bibnamefont {Ashoori}}, \bibinfo {author} {\bibfnamefont
  {N.}~\bibnamefont {Gedik}}, \bibinfo {author} {\bibfnamefont
  {L.}~\bibnamefont {Fu}}, \bibinfo {author} {\bibfnamefont {S.-Y.}\
  \bibnamefont {Xu}},\ and\ \bibinfo {author} {\bibfnamefont {P.}~\bibnamefont
  {Jarillo-Herrero}},\ }\bibfield  {title} {\bibinfo {title} {Unconventional
  ferroelectricity in moir{\'{e}} heterostructures},\ }\href
  {https://doi.org/10.1038/s41586-020-2970-9} {\bibfield  {journal} {\bibinfo
  {journal} {Nature}\ }\textbf {\bibinfo {volume} {588}},\ \bibinfo {pages}
  {71} (\bibinfo {year} {2020})}\BibitemShut {NoStop}%
\bibitem [{\citenamefont {Stern}\ \emph {et~al.}(2021)\citenamefont {Stern},
  \citenamefont {Waschitz}, \citenamefont {Cao}, \citenamefont {Nevo},
  \citenamefont {Watanabe}, \citenamefont {Taniguchi}, \citenamefont {Sela},
  \citenamefont {Urbakh}, \citenamefont {Hod},\ and\ \citenamefont
  {Shalom}}]{Stern21p1462}%
  \BibitemOpen
  \bibfield  {author} {\bibinfo {author} {\bibfnamefont {M.~V.}\ \bibnamefont
  {Stern}}, \bibinfo {author} {\bibfnamefont {Y.}~\bibnamefont {Waschitz}},
  \bibinfo {author} {\bibfnamefont {W.}~\bibnamefont {Cao}}, \bibinfo {author}
  {\bibfnamefont {I.}~\bibnamefont {Nevo}}, \bibinfo {author} {\bibfnamefont
  {K.}~\bibnamefont {Watanabe}}, \bibinfo {author} {\bibfnamefont
  {T.}~\bibnamefont {Taniguchi}}, \bibinfo {author} {\bibfnamefont
  {E.}~\bibnamefont {Sela}}, \bibinfo {author} {\bibfnamefont {M.}~\bibnamefont
  {Urbakh}}, \bibinfo {author} {\bibfnamefont {O.}~\bibnamefont {Hod}},\ and\
  \bibinfo {author} {\bibfnamefont {M.~B.}\ \bibnamefont {Shalom}},\ }\bibfield
   {title} {\bibinfo {title} {Interfacial ferroelectricity by van der {Waals}
  sliding},\ }\href {https://doi.org/10.1126/science.abe8177} {\bibfield
  {journal} {\bibinfo  {journal} {Science}\ }\textbf {\bibinfo {volume}
  {372}},\ \bibinfo {pages} {1462} (\bibinfo {year} {2021})}\BibitemShut
  {NoStop}%
\bibitem [{\citenamefont {Rog{\'{e}}e}\ \emph {et~al.}(2022)\citenamefont
  {Rog{\'{e}}e}, \citenamefont {Wang}, \citenamefont {Zhang}, \citenamefont
  {Cai}, \citenamefont {Wang}, \citenamefont {Chhowalla}, \citenamefont {Ji},\
  and\ \citenamefont {Lau}}]{Rogee22p973}%
  \BibitemOpen
  \bibfield  {author} {\bibinfo {author} {\bibfnamefont {L.}~\bibnamefont
  {Rog{\'{e}}e}}, \bibinfo {author} {\bibfnamefont {L.}~\bibnamefont {Wang}},
  \bibinfo {author} {\bibfnamefont {Y.}~\bibnamefont {Zhang}}, \bibinfo
  {author} {\bibfnamefont {S.}~\bibnamefont {Cai}}, \bibinfo {author}
  {\bibfnamefont {P.}~\bibnamefont {Wang}}, \bibinfo {author} {\bibfnamefont
  {M.}~\bibnamefont {Chhowalla}}, \bibinfo {author} {\bibfnamefont
  {W.}~\bibnamefont {Ji}},\ and\ \bibinfo {author} {\bibfnamefont {S.~P.}\
  \bibnamefont {Lau}},\ }\bibfield  {title} {\bibinfo {title} {Ferroelectricity
  in untwisted heterobilayers of transition metal dichalcogenides},\ }\href
  {https://doi.org/10.1126/science.abm5734} {\bibfield  {journal} {\bibinfo
  {journal} {Science}\ }\textbf {\bibinfo {volume} {376}},\ \bibinfo {pages}
  {973} (\bibinfo {year} {2022})}\BibitemShut {NoStop}%
\bibitem [{\citenamefont {Wu}\ and\ \citenamefont
  {Li}(2021)}]{Wu21pe2115703118}%
  \BibitemOpen
  \bibfield  {author} {\bibinfo {author} {\bibfnamefont {M.}~\bibnamefont
  {Wu}}\ and\ \bibinfo {author} {\bibfnamefont {J.}~\bibnamefont {Li}},\
  }\bibfield  {title} {\bibinfo {title} {Sliding ferroelectricity in 2d van der
  waals materials: Related physics and future opportunities},\ }\href@noop {}
  {\bibfield  {journal} {\bibinfo  {journal} {Proc. Natl. Acad. Sci.}\ }\textbf
  {\bibinfo {volume} {118}},\ \bibinfo {pages} {e2115703118} (\bibinfo {year}
  {2021})}\BibitemShut {NoStop}%
\bibitem [{\citenamefont {Miao}\ \emph {et~al.}(2022)\citenamefont {Miao},
  \citenamefont {Ding}, \citenamefont {Wang}, \citenamefont {Shi},
  \citenamefont {Ye}, \citenamefont {Li}, \citenamefont {Yao}, \citenamefont
  {Dong},\ and\ \citenamefont {Zhang}}]{Miao22p1158}%
  \BibitemOpen
  \bibfield  {author} {\bibinfo {author} {\bibfnamefont {L.-P.}\ \bibnamefont
  {Miao}}, \bibinfo {author} {\bibfnamefont {N.}~\bibnamefont {Ding}}, \bibinfo
  {author} {\bibfnamefont {N.}~\bibnamefont {Wang}}, \bibinfo {author}
  {\bibfnamefont {C.}~\bibnamefont {Shi}}, \bibinfo {author} {\bibfnamefont
  {H.-Y.}\ \bibnamefont {Ye}}, \bibinfo {author} {\bibfnamefont
  {L.}~\bibnamefont {Li}}, \bibinfo {author} {\bibfnamefont {Y.-F.}\
  \bibnamefont {Yao}}, \bibinfo {author} {\bibfnamefont {S.}~\bibnamefont
  {Dong}},\ and\ \bibinfo {author} {\bibfnamefont {Y.}~\bibnamefont {Zhang}},\
  }\bibfield  {title} {\bibinfo {title} {Direct observation of geometric and
  sliding ferroelectricity in an amphidynamic crystal},\ }\href
  {https://doi.org/10.1038/s41563-022-01322-1} {\bibfield  {journal} {\bibinfo
  {journal} {Nat. Mater.}\ }\textbf {\bibinfo {volume} {21}},\ \bibinfo {pages}
  {1158} (\bibinfo {year} {2022})}\BibitemShut {NoStop}%
\bibitem [{\citenamefont {Ielmini}\ and\ \citenamefont
  {Wong}(2018)}]{Ielmini18p333}%
  \BibitemOpen
  \bibfield  {author} {\bibinfo {author} {\bibfnamefont {D.}~\bibnamefont
  {Ielmini}}\ and\ \bibinfo {author} {\bibfnamefont {H.-S.~P.}\ \bibnamefont
  {Wong}},\ }\bibfield  {title} {\bibinfo {title} {In-memory computing with
  resistive switching devices},\ }\href
  {https://doi.org/10.1038/s41928-018-0092-2} {\bibfield  {journal} {\bibinfo
  {journal} {Nat. Electron.}\ }\textbf {\bibinfo {volume} {1}},\ \bibinfo
  {pages} {333} (\bibinfo {year} {2018})}\BibitemShut {NoStop}%
\bibitem [{\citenamefont {Yu}\ \emph {et~al.}(2017)\citenamefont {Yu},
  \citenamefont {Chung}, \citenamefont {Shewmon}, \citenamefont {Ho},
  \citenamefont {Carpenter}, \citenamefont {Larrabee}, \citenamefont {Sun},
  \citenamefont {Jones}, \citenamefont {Ade}, \citenamefont
  {O{\textquotesingle}Connor},\ and\ \citenamefont {So}}]{Yu17p1700461}%
  \BibitemOpen
  \bibfield  {author} {\bibinfo {author} {\bibfnamefont {H.}~\bibnamefont
  {Yu}}, \bibinfo {author} {\bibfnamefont {C.-C.}\ \bibnamefont {Chung}},
  \bibinfo {author} {\bibfnamefont {N.}~\bibnamefont {Shewmon}}, \bibinfo
  {author} {\bibfnamefont {S.}~\bibnamefont {Ho}}, \bibinfo {author}
  {\bibfnamefont {J.~H.}\ \bibnamefont {Carpenter}}, \bibinfo {author}
  {\bibfnamefont {R.}~\bibnamefont {Larrabee}}, \bibinfo {author}
  {\bibfnamefont {T.}~\bibnamefont {Sun}}, \bibinfo {author} {\bibfnamefont
  {J.~L.}\ \bibnamefont {Jones}}, \bibinfo {author} {\bibfnamefont
  {H.}~\bibnamefont {Ade}}, \bibinfo {author} {\bibfnamefont {B.~T.}\
  \bibnamefont {O{\textquotesingle}Connor}},\ and\ \bibinfo {author}
  {\bibfnamefont {F.}~\bibnamefont {So}},\ }\bibfield  {title} {\bibinfo
  {title} {Flexible inorganic ferroelectric thin films for nonvolatile memory
  devices},\ }\href {https://doi.org/10.1002/adfm.201700461} {\bibfield
  {journal} {\bibinfo  {journal} {Adv. Funct. Mater}\ }\textbf {\bibinfo
  {volume} {27}},\ \bibinfo {pages} {1700461} (\bibinfo {year}
  {2017})}\BibitemShut {NoStop}%
\bibitem [{\citenamefont {Si}\ \emph {et~al.}(2019)\citenamefont {Si},
  \citenamefont {Saha}, \citenamefont {Gao}, \citenamefont {Qiu}, \citenamefont
  {Qin}, \citenamefont {Duan}, \citenamefont {Jian}, \citenamefont {Niu},
  \citenamefont {Wang}, \citenamefont {Wu}, \citenamefont {Gupta},\ and\
  \citenamefont {Ye}}]{Si19p580}%
  \BibitemOpen
  \bibfield  {author} {\bibinfo {author} {\bibfnamefont {M.}~\bibnamefont
  {Si}}, \bibinfo {author} {\bibfnamefont {A.~K.}\ \bibnamefont {Saha}},
  \bibinfo {author} {\bibfnamefont {S.}~\bibnamefont {Gao}}, \bibinfo {author}
  {\bibfnamefont {G.}~\bibnamefont {Qiu}}, \bibinfo {author} {\bibfnamefont
  {J.}~\bibnamefont {Qin}}, \bibinfo {author} {\bibfnamefont {Y.}~\bibnamefont
  {Duan}}, \bibinfo {author} {\bibfnamefont {J.}~\bibnamefont {Jian}}, \bibinfo
  {author} {\bibfnamefont {C.}~\bibnamefont {Niu}}, \bibinfo {author}
  {\bibfnamefont {H.}~\bibnamefont {Wang}}, \bibinfo {author} {\bibfnamefont
  {W.}~\bibnamefont {Wu}}, \bibinfo {author} {\bibfnamefont {S.~K.}\
  \bibnamefont {Gupta}},\ and\ \bibinfo {author} {\bibfnamefont {P.~D.}\
  \bibnamefont {Ye}},\ }\bibfield  {title} {\bibinfo {title} {A ferroelectric
  semiconductor field-effect transistor},\ }\href
  {https://doi.org/10.1038/s41928-019-0338-7} {\bibfield  {journal} {\bibinfo
  {journal} {Nat. Electron}\ }\textbf {\bibinfo {volume} {2}},\ \bibinfo
  {pages} {580} (\bibinfo {year} {2019})}\BibitemShut {NoStop}%
\bibitem [{\citenamefont {Zhang}\ \emph {et~al.}(2021)\citenamefont {Zhang},
  \citenamefont {Wang}, \citenamefont {Chen}, \citenamefont {Ma}, \citenamefont
  {Lu},\ and\ \citenamefont {Loh}}]{Zhang21p2100609}%
  \BibitemOpen
  \bibfield  {author} {\bibinfo {author} {\bibfnamefont {Y.}~\bibnamefont
  {Zhang}}, \bibinfo {author} {\bibfnamefont {L.}~\bibnamefont {Wang}},
  \bibinfo {author} {\bibfnamefont {H.}~\bibnamefont {Chen}}, \bibinfo {author}
  {\bibfnamefont {T.}~\bibnamefont {Ma}}, \bibinfo {author} {\bibfnamefont
  {X.}~\bibnamefont {Lu}},\ and\ \bibinfo {author} {\bibfnamefont {K.~P.}\
  \bibnamefont {Loh}},\ }\bibfield  {title} {\bibinfo {title} {Analog and
  digital mode $\alpha$-{In}$_2${Se}$_3$ memristive devices for neuromorphic
  and memory applications},\ }\href {https://doi.org/10.1002/aelm.202100609}
  {\bibfield  {journal} {\bibinfo  {journal} {Adv. Electron. Mater}\ }\textbf
  {\bibinfo {volume} {7}},\ \bibinfo {pages} {2100609} (\bibinfo {year}
  {2021})}\BibitemShut {NoStop}%
\bibitem [{\citenamefont {Wan}\ \emph {et~al.}(2019)\citenamefont {Wan},
  \citenamefont {Li}, \citenamefont {Li}, \citenamefont {Mao}, \citenamefont
  {Wang}, \citenamefont {Chen}, \citenamefont {Dong}, \citenamefont {Nie},
  \citenamefont {Xiang}, \citenamefont {Liu}, \citenamefont {Zhu},\ and\
  \citenamefont {Zeng}}]{Wan19p1808606}%
  \BibitemOpen
  \bibfield  {author} {\bibinfo {author} {\bibfnamefont {S.}~\bibnamefont
  {Wan}}, \bibinfo {author} {\bibfnamefont {Y.}~\bibnamefont {Li}}, \bibinfo
  {author} {\bibfnamefont {W.}~\bibnamefont {Li}}, \bibinfo {author}
  {\bibfnamefont {X.}~\bibnamefont {Mao}}, \bibinfo {author} {\bibfnamefont
  {C.}~\bibnamefont {Wang}}, \bibinfo {author} {\bibfnamefont {C.}~\bibnamefont
  {Chen}}, \bibinfo {author} {\bibfnamefont {J.}~\bibnamefont {Dong}}, \bibinfo
  {author} {\bibfnamefont {A.}~\bibnamefont {Nie}}, \bibinfo {author}
  {\bibfnamefont {J.}~\bibnamefont {Xiang}}, \bibinfo {author} {\bibfnamefont
  {Z.}~\bibnamefont {Liu}}, \bibinfo {author} {\bibfnamefont {W.}~\bibnamefont
  {Zhu}},\ and\ \bibinfo {author} {\bibfnamefont {H.}~\bibnamefont {Zeng}},\
  }\bibfield  {title} {\bibinfo {title} {Nonvolatile ferroelectric memory
  effect in ultrathin $\alpha$-{In$_2$Se$_3$}},\ }\href
  {https://doi.org/10.1002/adfm.201808606} {\bibfield  {journal} {\bibinfo
  {journal} {Adv. Funct. Mater.}\ }\textbf {\bibinfo {volume} {29}},\ \bibinfo
  {pages} {1808606} (\bibinfo {year} {2019})}\BibitemShut {NoStop}%
\bibitem [{\citenamefont {Garcia}\ and\ \citenamefont
  {Bibes}(2014)}]{Garcia14p4289}%
  \BibitemOpen
  \bibfield  {author} {\bibinfo {author} {\bibfnamefont {V.}~\bibnamefont
  {Garcia}}\ and\ \bibinfo {author} {\bibfnamefont {M.}~\bibnamefont {Bibes}},\
  }\bibfield  {title} {\bibinfo {title} {Ferroelectric tunnel junctions for
  information storage and processing},\ }\href
  {https://doi.org/10.1038/ncomms5289} {\bibfield  {journal} {\bibinfo
  {journal} {Nat. Commun.}\ }\textbf {\bibinfo {volume} {5}},\ \bibinfo {pages}
  {4289} (\bibinfo {year} {2014})}\BibitemShut {NoStop}%
\bibitem [{\citenamefont {Junquera}\ and\ \citenamefont
  {Ghosez}(2003)}]{Junquera03p506}%
  \BibitemOpen
  \bibfield  {author} {\bibinfo {author} {\bibfnamefont {J.}~\bibnamefont
  {Junquera}}\ and\ \bibinfo {author} {\bibfnamefont {P.}~\bibnamefont
  {Ghosez}},\ }\bibfield  {title} {\bibinfo {title} {Critical thickness for
  ferroelectricity in perovskite ultrathin films},\ }\href
  {https://doi.org/10.1038/nature01501} {\bibfield  {journal} {\bibinfo
  {journal} {Nature}\ }\textbf {\bibinfo {volume} {422}},\ \bibinfo {pages}
  {506} (\bibinfo {year} {2003})}\BibitemShut {NoStop}%
\bibitem [{\citenamefont {Xue}\ \emph {et~al.}(2021)\citenamefont {Xue},
  \citenamefont {He}, \citenamefont {Ma}, \citenamefont {Zheng}, \citenamefont
  {Zhang}, \citenamefont {Li}, \citenamefont {He}, \citenamefont {Yu},\ and\
  \citenamefont {Zhang}}]{Xue21p7291}%
  \BibitemOpen
  \bibfield  {author} {\bibinfo {author} {\bibfnamefont {F.}~\bibnamefont
  {Xue}}, \bibinfo {author} {\bibfnamefont {X.}~\bibnamefont {He}}, \bibinfo
  {author} {\bibfnamefont {Y.}~\bibnamefont {Ma}}, \bibinfo {author}
  {\bibfnamefont {D.}~\bibnamefont {Zheng}}, \bibinfo {author} {\bibfnamefont
  {C.}~\bibnamefont {Zhang}}, \bibinfo {author} {\bibfnamefont {L.-J.}\
  \bibnamefont {Li}}, \bibinfo {author} {\bibfnamefont {J.-H.}\ \bibnamefont
  {He}}, \bibinfo {author} {\bibfnamefont {B.}~\bibnamefont {Yu}},\ and\
  \bibinfo {author} {\bibfnamefont {X.}~\bibnamefont {Zhang}},\ }\bibfield
  {title} {\bibinfo {title} {Unraveling the origin of ferroelectric resistance
  switching through the interfacial engineering of layered ferroelectric-metal
  junctions},\ }\href {https://doi.org/10.1038/s41467-021-27617-6} {\bibfield
  {journal} {\bibinfo  {journal} {Nat. Commun}\ }\textbf {\bibinfo {volume}
  {12}},\ \bibinfo {pages} {7291} (\bibinfo {year} {2021})}\BibitemShut
  {NoStop}%
\bibitem [{\citenamefont {Gerra}\ \emph {et~al.}(2006)\citenamefont {Gerra},
  \citenamefont {Tagantsev}, \citenamefont {Setter},\ and\ \citenamefont
  {Parlinski}}]{Gerra06p107603}%
  \BibitemOpen
  \bibfield  {author} {\bibinfo {author} {\bibfnamefont {G.}~\bibnamefont
  {Gerra}}, \bibinfo {author} {\bibfnamefont {A.~K.}\ \bibnamefont
  {Tagantsev}}, \bibinfo {author} {\bibfnamefont {N.}~\bibnamefont {Setter}},\
  and\ \bibinfo {author} {\bibfnamefont {K.}~\bibnamefont {Parlinski}},\
  }\bibfield  {title} {\bibinfo {title} {Ionic polarizability of conductive
  metal oxides and critical thickness for ferroelectricity in {BaTiO$_3$}},\
  }\href@noop {} {\bibfield  {journal} {\bibinfo  {journal} {Phys. Rev. Lett.}\
  }\textbf {\bibinfo {volume} {96}},\ \bibinfo {pages} {107603} (\bibinfo
  {year} {2006})}\BibitemShut {NoStop}%
\bibitem [{\citenamefont {Yang}\ \emph {et~al.}(2022)\citenamefont {Yang},
  \citenamefont {Zhou}, \citenamefont {Lu}, \citenamefont {Luo}, \citenamefont
  {Yang},\ and\ \citenamefont {Shen}}]{Yang22p1422}%
  \BibitemOpen
  \bibfield  {author} {\bibinfo {author} {\bibfnamefont {J.}~\bibnamefont
  {Yang}}, \bibinfo {author} {\bibfnamefont {J.}~\bibnamefont {Zhou}}, \bibinfo
  {author} {\bibfnamefont {J.}~\bibnamefont {Lu}}, \bibinfo {author}
  {\bibfnamefont {Z.}~\bibnamefont {Luo}}, \bibinfo {author} {\bibfnamefont
  {J.}~\bibnamefont {Yang}},\ and\ \bibinfo {author} {\bibfnamefont
  {L.}~\bibnamefont {Shen}},\ }\bibfield  {title} {\bibinfo {title} {Giant
  tunnelling electroresistance through {2D} sliding ferroelectric materials},\
  }\href {https://doi.org/10.1039/d2mh00080f} {\bibfield  {journal} {\bibinfo
  {journal} {Mater. Horiz.}\ }\textbf {\bibinfo {volume} {9}},\ \bibinfo
  {pages} {1422–1430} (\bibinfo {year} {2022})}\BibitemShut {NoStop}%
\bibitem [{\citenamefont {Wang}\ \emph {et~al.}(2012)\citenamefont {Wang},
  \citenamefont {Liu}, \citenamefont {Burton}, \citenamefont {Jaswal},\ and\
  \citenamefont {Tsymbal}}]{Wang12p247601}%
  \BibitemOpen
  \bibfield  {author} {\bibinfo {author} {\bibfnamefont {Y.}~\bibnamefont
  {Wang}}, \bibinfo {author} {\bibfnamefont {X.}~\bibnamefont {Liu}}, \bibinfo
  {author} {\bibfnamefont {J.~D.}\ \bibnamefont {Burton}}, \bibinfo {author}
  {\bibfnamefont {S.~S.}\ \bibnamefont {Jaswal}},\ and\ \bibinfo {author}
  {\bibfnamefont {E.~Y.}\ \bibnamefont {Tsymbal}},\ }\bibfield  {title}
  {\bibinfo {title} {Ferroelectric instability under screened coulomb
  interactions},\ }\href {https://doi.org/10.1103/PhysRevLett.109.247601}
  {\bibfield  {journal} {\bibinfo  {journal} {Phys. Rev. Lett.}\ }\textbf
  {\bibinfo {volume} {109}},\ \bibinfo {pages} {247601} (\bibinfo {year}
  {2012})}\BibitemShut {NoStop}%
\bibitem [{\citenamefont {Zhang}\ \emph {et~al.}(2017)\citenamefont {Zhang},
  \citenamefont {Zhu}, \citenamefont {Tang}, \citenamefont {Liu}, \citenamefont
  {Li}, \citenamefont {Han}, \citenamefont {Ma}, \citenamefont {Wu},
  \citenamefont {Chen}, \citenamefont {Saremi},\ and\ \citenamefont
  {Ma}}]{Zhang17p1703543}%
  \BibitemOpen
  \bibfield  {author} {\bibinfo {author} {\bibfnamefont {S.}~\bibnamefont
  {Zhang}}, \bibinfo {author} {\bibfnamefont {Y.}~\bibnamefont {Zhu}}, \bibinfo
  {author} {\bibfnamefont {Y.}~\bibnamefont {Tang}}, \bibinfo {author}
  {\bibfnamefont {Y.}~\bibnamefont {Liu}}, \bibinfo {author} {\bibfnamefont
  {S.}~\bibnamefont {Li}}, \bibinfo {author} {\bibfnamefont {M.}~\bibnamefont
  {Han}}, \bibinfo {author} {\bibfnamefont {J.}~\bibnamefont {Ma}}, \bibinfo
  {author} {\bibfnamefont {B.}~\bibnamefont {Wu}}, \bibinfo {author}
  {\bibfnamefont {Z.}~\bibnamefont {Chen}}, \bibinfo {author} {\bibfnamefont
  {S.}~\bibnamefont {Saremi}},\ and\ \bibinfo {author} {\bibfnamefont
  {X.}~\bibnamefont {Ma}},\ }\bibfield  {title} {\bibinfo {title} {Giant
  polarization sustainability in ultrathin ferroelectric films stabilized by
  charge transfer},\ }\href {https://doi.org/10.1002/adma.201703543} {\bibfield
   {journal} {\bibinfo  {journal} {Adv. Mater.}\ }\textbf {\bibinfo {volume}
  {29}},\ \bibinfo {pages} {1703543} (\bibinfo {year} {2017})}\BibitemShut
  {NoStop}%
\bibitem [{\citenamefont {Lin}\ \emph {et~al.}(2021)\citenamefont {Lin},
  \citenamefont {Shao}, \citenamefont {Dai}, \citenamefont {Li}, \citenamefont
  {Liu}, \citenamefont {Dai}, \citenamefont {Xiao}, \citenamefont {Deng},
  \citenamefont {Gruverman}, \citenamefont {Zeng},\ and\ \citenamefont
  {Huang}}]{Lin21p7}%
  \BibitemOpen
  \bibfield  {author} {\bibinfo {author} {\bibfnamefont {Y.}~\bibnamefont
  {Lin}}, \bibinfo {author} {\bibfnamefont {Y.}~\bibnamefont {Shao}}, \bibinfo
  {author} {\bibfnamefont {J.}~\bibnamefont {Dai}}, \bibinfo {author}
  {\bibfnamefont {T.}~\bibnamefont {Li}}, \bibinfo {author} {\bibfnamefont
  {Y.}~\bibnamefont {Liu}}, \bibinfo {author} {\bibfnamefont {X.}~\bibnamefont
  {Dai}}, \bibinfo {author} {\bibfnamefont {X.}~\bibnamefont {Xiao}}, \bibinfo
  {author} {\bibfnamefont {Y.}~\bibnamefont {Deng}}, \bibinfo {author}
  {\bibfnamefont {A.}~\bibnamefont {Gruverman}}, \bibinfo {author}
  {\bibfnamefont {X.~C.}\ \bibnamefont {Zeng}},\ and\ \bibinfo {author}
  {\bibfnamefont {J.}~\bibnamefont {Huang}},\ }\bibfield  {title} {\bibinfo
  {title} {Metallic surface doping of metal halide perovskites},\ }\href
  {https://doi.org/10.1038/s41467-020-20110-6} {\bibfield  {journal} {\bibinfo
  {journal} {Nat. Commun.}\ }\textbf {\bibinfo {volume} {12}},\ \bibinfo
  {pages} {7} (\bibinfo {year} {2021})}\BibitemShut {NoStop}%
\bibitem [{\citenamefont {Wu}\ \emph {et~al.}(2021)\citenamefont {Wu},
  \citenamefont {Bai}, \citenamefont {Huang}, \citenamefont {Ma}, \citenamefont
  {Liu},\ and\ \citenamefont {Liu}}]{Wu21p174107}%
  \BibitemOpen
  \bibfield  {author} {\bibinfo {author} {\bibfnamefont {J.}~\bibnamefont
  {Wu}}, \bibinfo {author} {\bibfnamefont {L.}~\bibnamefont {Bai}}, \bibinfo
  {author} {\bibfnamefont {J.}~\bibnamefont {Huang}}, \bibinfo {author}
  {\bibfnamefont {L.}~\bibnamefont {Ma}}, \bibinfo {author} {\bibfnamefont
  {J.}~\bibnamefont {Liu}},\ and\ \bibinfo {author} {\bibfnamefont
  {S.}~\bibnamefont {Liu}},\ }\bibfield  {title} {\bibinfo {title} {Accurate
  force field of two-dimensional ferroelectrics from deep learning},\ }\href
  {https://doi.org/10.1103/physrevb.104.174107} {\bibfield  {journal} {\bibinfo
   {journal} {Phys. Rev. B}\ }\textbf {\bibinfo {volume} {104}},\ \bibinfo
  {pages} {174107} (\bibinfo {year} {2021})}\BibitemShut {NoStop}%
\bibitem [{\citenamefont {Liu}\ \emph {et~al.}(2018)\citenamefont {Liu},
  \citenamefont {Lu}, \citenamefont {Picozzi}, \citenamefont {Bellaiche},\ and\
  \citenamefont {Xiang}}]{Liu18p027601}%
  \BibitemOpen
  \bibfield  {author} {\bibinfo {author} {\bibfnamefont {K.}~\bibnamefont
  {Liu}}, \bibinfo {author} {\bibfnamefont {J.}~\bibnamefont {Lu}}, \bibinfo
  {author} {\bibfnamefont {S.}~\bibnamefont {Picozzi}}, \bibinfo {author}
  {\bibfnamefont {L.}~\bibnamefont {Bellaiche}},\ and\ \bibinfo {author}
  {\bibfnamefont {H.}~\bibnamefont {Xiang}},\ }\bibfield  {title} {\bibinfo
  {title} {Intrinsic origin of enhancement of ferroelectricity in {SnTe}
  ultrathin films},\ }\href {https://doi.org/10.1103/physrevlett.121.027601}
  {\bibfield  {journal} {\bibinfo  {journal} {Phys. Rev. Lett.}\ }\textbf
  {\bibinfo {volume} {121}},\ \bibinfo {pages} {027601} (\bibinfo {year}
  {2018})}\BibitemShut {NoStop}%
\bibitem [{\citenamefont {Kresse}\ and\ \citenamefont
  {J}(1996{\natexlab{a}})}]{Kresse96p11169}%
  \BibitemOpen
  \bibfield  {author} {\bibinfo {author} {\bibfnamefont {G.}~\bibnamefont
  {Kresse}}\ and\ \bibinfo {author} {\bibfnamefont {F.}~\bibnamefont {J}},\
  }\bibfield  {title} {\bibinfo {title} {Efficient iterative schemes for $ab$
  $initio$ total-energy calculations using a plane-wave basis set},\
  }\href@noop {} {\bibfield  {journal} {\bibinfo  {journal} {Phys. Rev. B}\
  }\textbf {\bibinfo {volume} {54}},\ \bibinfo {pages} {11169} (\bibinfo {year}
  {1996}{\natexlab{a}})}\BibitemShut {NoStop}%
\bibitem [{\citenamefont {Kresse}\ and\ \citenamefont
  {J}(1996{\natexlab{b}})}]{Kresse96p15}%
  \BibitemOpen
  \bibfield  {author} {\bibinfo {author} {\bibfnamefont {G.}~\bibnamefont
  {Kresse}}\ and\ \bibinfo {author} {\bibfnamefont {F.}~\bibnamefont {J}},\
  }\bibfield  {title} {\bibinfo {title} {Efficiency of ab-initio total energy
  calculations for metals and semiconductors using a plane-wave basis set},\
  }\href@noop {} {\bibfield  {journal} {\bibinfo  {journal} {Comput. Mater.
  Sci.}\ }\textbf {\bibinfo {volume} {6}},\ \bibinfo {pages} {15} (\bibinfo
  {year} {1996}{\natexlab{b}})}\BibitemShut {NoStop}%
\bibitem [{\citenamefont {Perdew}\ \emph {et~al.}(1996)\citenamefont {Perdew},
  \citenamefont {Burke},\ and\ \citenamefont {Ernzerhof}}]{Perdew96p3865}%
  \BibitemOpen
  \bibfield  {author} {\bibinfo {author} {\bibfnamefont {J.~P.}\ \bibnamefont
  {Perdew}}, \bibinfo {author} {\bibfnamefont {K.}~\bibnamefont {Burke}},\ and\
  \bibinfo {author} {\bibfnamefont {M.}~\bibnamefont {Ernzerhof}},\ }\bibfield
  {title} {\bibinfo {title} {Generalized gradient approximation made simple},\
  }\href {https://doi.org/10.1103/PhysRevLett.77.3865} {\bibfield  {journal}
  {\bibinfo  {journal} {Phys. Rev. Lett.}\ }\textbf {\bibinfo {volume} {77}},\
  \bibinfo {pages} {3865} (\bibinfo {year} {1996})}\BibitemShut {NoStop}%
\bibitem [{\citenamefont {Blochl}(1994)}]{Blochi94p17953}%
  \BibitemOpen
  \bibfield  {author} {\bibinfo {author} {\bibfnamefont {P.~E.}\ \bibnamefont
  {Blochl}},\ }\bibfield  {title} {\bibinfo {title} {Projector augmented-wave
  method},\ }\href@noop {} {\bibfield  {journal} {\bibinfo  {journal} {Phys.
  Rev. B}\ }\textbf {\bibinfo {volume} {50}},\ \bibinfo {pages} {17953}
  (\bibinfo {year} {1994})}\BibitemShut {NoStop}%
\bibitem [{\citenamefont {Haastrup}\ \emph {et~al.}(2018)\citenamefont
  {Haastrup}, \citenamefont {Strange}, \citenamefont {Pandey}, \citenamefont
  {Deilmann}, \citenamefont {Schmidt}, \citenamefont {Hinsche}, \citenamefont
  {Gjerding}, \citenamefont {Torelli}, \citenamefont {Larsen}, \citenamefont
  {Riis-Jensen}, \citenamefont {Gath}, \citenamefont {Jacobsen}, \citenamefont
  {Mortensen}, \citenamefont {Olsen},\ and\ \citenamefont
  {Thygesen}}]{Haastrup18p042002}%
  \BibitemOpen
  \bibfield  {author} {\bibinfo {author} {\bibfnamefont {S.}~\bibnamefont
  {Haastrup}}, \bibinfo {author} {\bibfnamefont {M.}~\bibnamefont {Strange}},
  \bibinfo {author} {\bibfnamefont {M.}~\bibnamefont {Pandey}}, \bibinfo
  {author} {\bibfnamefont {T.}~\bibnamefont {Deilmann}}, \bibinfo {author}
  {\bibfnamefont {P.~S.}\ \bibnamefont {Schmidt}}, \bibinfo {author}
  {\bibfnamefont {N.~F.}\ \bibnamefont {Hinsche}}, \bibinfo {author}
  {\bibfnamefont {M.~N.}\ \bibnamefont {Gjerding}}, \bibinfo {author}
  {\bibfnamefont {D.}~\bibnamefont {Torelli}}, \bibinfo {author} {\bibfnamefont
  {P.~M.}\ \bibnamefont {Larsen}}, \bibinfo {author} {\bibfnamefont {A.~C.}\
  \bibnamefont {Riis-Jensen}}, \bibinfo {author} {\bibfnamefont
  {J.}~\bibnamefont {Gath}}, \bibinfo {author} {\bibfnamefont {K.~W.}\
  \bibnamefont {Jacobsen}}, \bibinfo {author} {\bibfnamefont {J.~J.}\
  \bibnamefont {Mortensen}}, \bibinfo {author} {\bibfnamefont {T.}~\bibnamefont
  {Olsen}},\ and\ \bibinfo {author} {\bibfnamefont {K.~S.}\ \bibnamefont
  {Thygesen}},\ }\bibfield  {title} {\bibinfo {title} {The computational {2D}
  materials database: high-throughput modeling and discovery of atomically thin
  crystals},\ }\href {https://doi.org/10.1088/2053-1583/aacfc1} {\bibfield
  {journal} {\bibinfo  {journal} {2D Mater.}\ }\textbf {\bibinfo {volume}
  {5}},\ \bibinfo {pages} {042002} (\bibinfo {year} {2018})}\BibitemShut
  {NoStop}%
\bibitem [{\citenamefont {Gjerding}\ \emph {et~al.}(2021)\citenamefont
  {Gjerding}, \citenamefont {Taghizadeh}, \citenamefont {Rasmussen},
  \citenamefont {Ali}, \citenamefont {Bertoldo}, \citenamefont {Deilmann},
  \citenamefont {Kn{\o}sgaard}, \citenamefont {Kruse}, \citenamefont {Larsen},
  \citenamefont {Manti}, \citenamefont {Pedersen}, \citenamefont {Petralanda},
  \citenamefont {Skovhus}, \citenamefont {Svendsen}, \citenamefont {Mortensen},
  \citenamefont {Olsen},\ and\ \citenamefont {Thygesen}}]{Gjerding21p044002}%
  \BibitemOpen
  \bibfield  {author} {\bibinfo {author} {\bibfnamefont {M.~N.}\ \bibnamefont
  {Gjerding}}, \bibinfo {author} {\bibfnamefont {A.}~\bibnamefont
  {Taghizadeh}}, \bibinfo {author} {\bibfnamefont {A.}~\bibnamefont
  {Rasmussen}}, \bibinfo {author} {\bibfnamefont {S.}~\bibnamefont {Ali}},
  \bibinfo {author} {\bibfnamefont {F.}~\bibnamefont {Bertoldo}}, \bibinfo
  {author} {\bibfnamefont {T.}~\bibnamefont {Deilmann}}, \bibinfo {author}
  {\bibfnamefont {N.~R.}\ \bibnamefont {Kn{\o}sgaard}}, \bibinfo {author}
  {\bibfnamefont {M.}~\bibnamefont {Kruse}}, \bibinfo {author} {\bibfnamefont
  {A.~H.}\ \bibnamefont {Larsen}}, \bibinfo {author} {\bibfnamefont
  {S.}~\bibnamefont {Manti}}, \bibinfo {author} {\bibfnamefont {T.~G.}\
  \bibnamefont {Pedersen}}, \bibinfo {author} {\bibfnamefont {U.}~\bibnamefont
  {Petralanda}}, \bibinfo {author} {\bibfnamefont {T.}~\bibnamefont {Skovhus}},
  \bibinfo {author} {\bibfnamefont {M.~K.}\ \bibnamefont {Svendsen}}, \bibinfo
  {author} {\bibfnamefont {J.~J.}\ \bibnamefont {Mortensen}}, \bibinfo {author}
  {\bibfnamefont {T.}~\bibnamefont {Olsen}},\ and\ \bibinfo {author}
  {\bibfnamefont {K.~S.}\ \bibnamefont {Thygesen}},\ }\bibfield  {title}
  {\bibinfo {title} {Recent progress of the computational {2D} materials
  database {(C2DB)}},\ }\href {https://doi.org/10.1088/2053-1583/ac1059}
  {\bibfield  {journal} {\bibinfo  {journal} {2D Mater.}\ }\textbf {\bibinfo
  {volume} {8}},\ \bibinfo {pages} {044002} (\bibinfo {year}
  {2021})}\BibitemShut {NoStop}%
\bibitem [{\citenamefont {Chen}\ \emph {et~al.}(2023)\citenamefont {Chen},
  \citenamefont {Sun}, \citenamefont {Li}, \citenamefont {Huang},\ and\
  \citenamefont {Loh}}]{Chen23p1077}%
  \BibitemOpen
  \bibfield  {author} {\bibinfo {author} {\bibfnamefont {Z.}~\bibnamefont
  {Chen}}, \bibinfo {author} {\bibfnamefont {M.}~\bibnamefont {Sun}}, \bibinfo
  {author} {\bibfnamefont {H.}~\bibnamefont {Li}}, \bibinfo {author}
  {\bibfnamefont {B.}~\bibnamefont {Huang}},\ and\ \bibinfo {author}
  {\bibfnamefont {K.~P.}\ \bibnamefont {Loh}},\ }\bibfield  {title} {\bibinfo
  {title} {Oscillatory order{\textendash}disorder transition during
  layer-by-layer growth of indium selenide},\ }\href
  {https://doi.org/10.1021/acs.nanolett.2c04785} {\bibfield  {journal}
  {\bibinfo  {journal} {Nano Lett.}\ }\textbf {\bibinfo {volume} {23}},\
  \bibinfo {pages} {1077} (\bibinfo {year} {2023})}\BibitemShut {NoStop}%
\bibitem [{\citenamefont {Zhang}\ \emph {et~al.}(2019)\citenamefont {Zhang},
  \citenamefont {Wang}, \citenamefont {Dong}, \citenamefont {Nie},
  \citenamefont {Xiang}, \citenamefont {Zhu}, \citenamefont {Liu},\ and\
  \citenamefont {Tao}}]{Zhang19p8004}%
  \BibitemOpen
  \bibfield  {author} {\bibinfo {author} {\bibfnamefont {F.}~\bibnamefont
  {Zhang}}, \bibinfo {author} {\bibfnamefont {Z.}~\bibnamefont {Wang}},
  \bibinfo {author} {\bibfnamefont {J.}~\bibnamefont {Dong}}, \bibinfo {author}
  {\bibfnamefont {A.}~\bibnamefont {Nie}}, \bibinfo {author} {\bibfnamefont
  {J.}~\bibnamefont {Xiang}}, \bibinfo {author} {\bibfnamefont
  {W.}~\bibnamefont {Zhu}}, \bibinfo {author} {\bibfnamefont {Z.}~\bibnamefont
  {Liu}},\ and\ \bibinfo {author} {\bibfnamefont {C.}~\bibnamefont {Tao}},\
  }\bibfield  {title} {\bibinfo {title} {Atomic-scale observation of reversible
  thermally driven phase transformation in 2{D} {In$_2$Se$_3$}},\ }\href
  {https://doi.org/10.1021/acsnano.9b02764} {\bibfield  {journal} {\bibinfo
  {journal} {{ACS} Nano}\ }\textbf {\bibinfo {volume} {13}},\ \bibinfo {pages}
  {8004} (\bibinfo {year} {2019})}\BibitemShut {NoStop}%
\bibitem [{\citenamefont {Zheng}\ \emph {et~al.}(2022)\citenamefont {Zheng},
  \citenamefont {Han}, \citenamefont {Yang}, \citenamefont {Wong},
  \citenamefont {Tsang}, \citenamefont {Lai}, \citenamefont {Zheng},
  \citenamefont {Yang}, \citenamefont {Lau}, \citenamefont {Ly}, \citenamefont
  {Yang},\ and\ \citenamefont {Zhao}}]{Zheng22peabo0773}%
  \BibitemOpen
  \bibfield  {author} {\bibinfo {author} {\bibfnamefont {X.}~\bibnamefont
  {Zheng}}, \bibinfo {author} {\bibfnamefont {W.}~\bibnamefont {Han}}, \bibinfo
  {author} {\bibfnamefont {K.}~\bibnamefont {Yang}}, \bibinfo {author}
  {\bibfnamefont {L.~W.}\ \bibnamefont {Wong}}, \bibinfo {author}
  {\bibfnamefont {C.~S.}\ \bibnamefont {Tsang}}, \bibinfo {author}
  {\bibfnamefont {K.~H.}\ \bibnamefont {Lai}}, \bibinfo {author} {\bibfnamefont
  {F.}~\bibnamefont {Zheng}}, \bibinfo {author} {\bibfnamefont
  {T.}~\bibnamefont {Yang}}, \bibinfo {author} {\bibfnamefont {S.~P.}\
  \bibnamefont {Lau}}, \bibinfo {author} {\bibfnamefont {T.~H.}\ \bibnamefont
  {Ly}}, \bibinfo {author} {\bibfnamefont {M.}~\bibnamefont {Yang}},\ and\
  \bibinfo {author} {\bibfnamefont {J.}~\bibnamefont {Zhao}},\ }\bibfield
  {title} {\bibinfo {title} {Phase and polarization modulation in
  two-dimensional {In$_2$Se$_3$} via in situ transmission electron
  microscopy},\ }\href {https://doi.org/10.1126/sciadv.abo0773} {\bibfield
  {journal} {\bibinfo  {journal} {Sci. Adv.}\ }\textbf {\bibinfo {volume}
  {8}},\ \bibinfo {pages} {eabo0773} (\bibinfo {year} {2022})}\BibitemShut
  {NoStop}%
\bibitem [{\citenamefont {Zheng}\ \emph {et~al.}(2018)\citenamefont {Zheng},
  \citenamefont {Yu}, \citenamefont {Zhu}, \citenamefont {Collins},
  \citenamefont {Kim}, \citenamefont {Lou}, \citenamefont {Xu}, \citenamefont
  {Li}, \citenamefont {Wei}, \citenamefont {Zhang}, \citenamefont {Edmonds},
  \citenamefont {Li}, \citenamefont {Seidel}, \citenamefont {Zhu},
  \citenamefont {Liu}, \citenamefont {Tang},\ and\ \citenamefont
  {Fuhrer}}]{Zheng18peaar7720}%
  \BibitemOpen
  \bibfield  {author} {\bibinfo {author} {\bibfnamefont {C.}~\bibnamefont
  {Zheng}}, \bibinfo {author} {\bibfnamefont {L.}~\bibnamefont {Yu}}, \bibinfo
  {author} {\bibfnamefont {L.}~\bibnamefont {Zhu}}, \bibinfo {author}
  {\bibfnamefont {J.~L.}\ \bibnamefont {Collins}}, \bibinfo {author}
  {\bibfnamefont {D.}~\bibnamefont {Kim}}, \bibinfo {author} {\bibfnamefont
  {Y.}~\bibnamefont {Lou}}, \bibinfo {author} {\bibfnamefont {C.}~\bibnamefont
  {Xu}}, \bibinfo {author} {\bibfnamefont {M.}~\bibnamefont {Li}}, \bibinfo
  {author} {\bibfnamefont {Z.}~\bibnamefont {Wei}}, \bibinfo {author}
  {\bibfnamefont {Y.}~\bibnamefont {Zhang}}, \bibinfo {author} {\bibfnamefont
  {M.~T.}\ \bibnamefont {Edmonds}}, \bibinfo {author} {\bibfnamefont
  {S.}~\bibnamefont {Li}}, \bibinfo {author} {\bibfnamefont {J.}~\bibnamefont
  {Seidel}}, \bibinfo {author} {\bibfnamefont {Y.}~\bibnamefont {Zhu}},
  \bibinfo {author} {\bibfnamefont {J.~Z.}\ \bibnamefont {Liu}}, \bibinfo
  {author} {\bibfnamefont {W.-X.}\ \bibnamefont {Tang}},\ and\ \bibinfo
  {author} {\bibfnamefont {M.~S.}\ \bibnamefont {Fuhrer}},\ }\bibfield  {title}
  {\bibinfo {title} {Room temperature in-plane ferroelectricity in van der
  {Waals} {In$_2$Se$_3$}},\ }\href {https://doi.org/10.1126/sciadv.aar7720}
  {\bibfield  {journal} {\bibinfo  {journal} {Sci. Adv.}\ }\textbf {\bibinfo
  {volume} {4}},\ \bibinfo {pages} {eaar7720} (\bibinfo {year}
  {2018})}\BibitemShut {NoStop}%
\bibitem [{\citenamefont {Huang}\ \emph {et~al.}(2022)\citenamefont {Huang},
  \citenamefont {Duan}, \citenamefont {Jeon}, \citenamefont {Kim},
  \citenamefont {Zhou}, \citenamefont {Li},\ and\ \citenamefont
  {Liu}}]{Huang22p1440}%
  \BibitemOpen
  \bibfield  {author} {\bibinfo {author} {\bibfnamefont {J.}~\bibnamefont
  {Huang}}, \bibinfo {author} {\bibfnamefont {X.}~\bibnamefont {Duan}},
  \bibinfo {author} {\bibfnamefont {S.}~\bibnamefont {Jeon}}, \bibinfo {author}
  {\bibfnamefont {Y.}~\bibnamefont {Kim}}, \bibinfo {author} {\bibfnamefont
  {J.}~\bibnamefont {Zhou}}, \bibinfo {author} {\bibfnamefont {J.}~\bibnamefont
  {Li}},\ and\ \bibinfo {author} {\bibfnamefont {S.}~\bibnamefont {Liu}},\
  }\bibfield  {title} {\bibinfo {title} {On-demand quantum spin hall insulators
  controlled by two-dimensional ferroelectricity},\ }\href
  {https://doi.org/10.1039/d2mh00334a} {\bibfield  {journal} {\bibinfo
  {journal} {Mater. Horiz.}\ }\textbf {\bibinfo {volume} {9}},\ \bibinfo
  {pages} {1440} (\bibinfo {year} {2022})}\BibitemShut {NoStop}%
\bibitem [{\citenamefont {Ke}\ \emph {et~al.}(2021)\citenamefont {Ke},
  \citenamefont {Huang},\ and\ \citenamefont {Liu}}]{Ke21p3387}%
  \BibitemOpen
  \bibfield  {author} {\bibinfo {author} {\bibfnamefont {C.}~\bibnamefont
  {Ke}}, \bibinfo {author} {\bibfnamefont {J.}~\bibnamefont {Huang}},\ and\
  \bibinfo {author} {\bibfnamefont {S.}~\bibnamefont {Liu}},\ }\bibfield
  {title} {\bibinfo {title} {Two-dimensional ferroelectric metal for
  electrocatalysis},\ }\href {https://doi.org/10.1039/d1mh01556g} {\bibfield
  {journal} {\bibinfo  {journal} {Mater. Horiz.}\ }\textbf {\bibinfo {volume}
  {8}},\ \bibinfo {pages} {3387} (\bibinfo {year} {2021})}\BibitemShut
  {NoStop}%
\end{thebibliography}%

\bigskip
\bigskip
\bigskip

\newpage
 
 \newpage

\clearpage
\newpage
 
\end{document}